\documentclass[10pt]{article}
\usepackage[ruled]{algorithm2e}
\renewcommand{\algorithmcfname}{ALGORITHM}
\SetAlFnt{\small}
\SetAlCapFnt{\small}
\SetAlCapNameFnt{\small}
\SetAlCapHSkip{0pt}
\IncMargin{-\parindent}

\usepackage{tikz}

\usepackage{amssymb, url}
\usepackage{amsmath,amsthm, xpatch}
\usepackage{mathtools}
\usepackage{dsfont}
\usepackage{verbatim}
\usepackage{graphicx}
\usepackage{epstopdf}
\usepackage{fullpage}
\usepackage[hang,flushmargin]{footmisc}
\usepackage{tikz}
\usepackage{enumitem, hyperref}
\usepackage{datetime}
\usepackage{bbm}
\usepackage{color}
\usetikzlibrary{calc}
\usetikzlibrary{decorations.pathreplacing,calligraphy}

\newtheorem{definition}{Definition}

\makeatletter
\newcommand{\newreptheorem}[2]{\newtheorem*{rep@#1}{\rep@title}\newenvironment{rep#1}[1]{\def\rep@title{#2 \ref*{##1}}\begin{rep@#1}}{\end{rep@#1}}}
\makeatother
\newreptheorem{theorem}{Theorem}

\theoremstyle{plain}

\newcounter{caseCount}
\AtBeginEnvironment{claim}{\setcounter{caseCount}{0}}

\providecommand{\keywords}[1]
{
  \small	
  \textbf{\textit{Keywords---}} #1
}

\begin{document}
\title{Matrix Scaling: a New Heuristic for the Feedback Vertex Set Problem}
\author{James M.\ Shook$^{1,2,3}$ and  Isabel Beichl$^{2}$}

\footnotetext[1]{National Institute of Standards and Technology, Computer Security Division, Gaithersburg, MD}
\footnotetext[2]{{\tt james.shook@nist.gov}}
\footnotetext[3]{This article is a U.S. Government work and is not subject to copyright in the USA.}

\date{Spring 2014}
\maketitle
\begin{abstract}For a digraph $G$, a set $F\subseteq V(G)$ is said to be a feedback vertex set (FVS) if $G-F$ is acyclic. The problem of finding a smallest FVS is NP-hard. We present a matrix scaling technique for finding feedback vertex sets in un-weighted directed graphs that runs in $O(|F|\log(|V|)|V|^{2})$ time. Our technique is empirically shown to produce smaller feedback vertex sets than other known heuristics and in a shorter amount of time.
\end{abstract}


\subsubsection*{Authors Note}
This is an unreleased draft we wrote in the Spring of 2014. However, we gave several presentations in 2014 and released the slides \cite{Shook2014}. In light of one of the entrants \cite{Bathie2022} to the PACE-2022 competition on the Directed Feedback Vertex Set (https://pacechallenge.org/2022/directed-fvs/) referencing our slides and method we are releasing the paper.

\section{Introduction}
\label{sec:Introduction}

We present a new heuristic for finding feedback vertex sets in unweighted directed graphs that is not only faster in the worst case than known heuristics but whose output quality is empirically better. Our idea is to slightly modify the adjacency matrix of a given digraph so that a small number of iterations of the Sinkhorn-Knopp algorithm (See Section~\ref{sec:sinkhorn-balancing}) produces a vertex that we can use to create a minimal feedback vertex set. Although our algorithm can be randomized, we limit our scope to deterministic methods so that we may briefly present the technique. 

We call a pair $G=(V,A)$ a directed graph (digraph) with vertices $V(G)$ and a set of ordered pairs of the vertices called arcs $A(G)$. A directed cycle of a digraph is an ordering $v_{0} \ldots v_{t-1}$ of vertices in $G$ such that for all $i<t-1$ the pairs $(v_{i},v_{i+1})$ and $(v_{t-1},v_{0})$ are arcs in $A(G)$. A digraph that does not have a directed cycle as a subgraph is said to be \textbf{acyclic}. We will use the abbreviation DAG to denote a directed acyclic digraph. A set $F\subseteq V(G)$ is said to be a \textbf{feedback vertex set}, denoted by FVS, if for any cycle $C$ in $G$ some vertex of $C$ is in $F$. An FVS is said to be \textbf{minimal} if no proper subset of $F$ is an FVS. For a digraph $G$, we are interested in finding a minimum FVS. We will denote the order of a minimum FVS of a digraph $G$ by $\tau(G)$. Clearly, if a vertex is incident with a loop, then that vertex is in every FVS, thus, we will proceed with the convention that our digraphs have no loops or parallel edges unless stated otherwise. Our notation follows close to that of \cite{DBLP:books/daglib/0030488}.

Finding feedback vertex sets has been useful in resolving deadlock in dependency digraphs. Two nodes in a dependency graph have an arc if one node is dependent on another in the system. As an example, consider two processes that share resources in an operating system. If the two processes cannot finish because they are waiting for each other to terminate, then the system has a deadlock. Killing one of the processes would allow the other to finish. This indicates that if we found a minimum FVS in a dependency digraph we could kill the fewest number of processes to clear the deadlock.

Unfortunately, the problem of finding an FVS of minimum size was shown by Karp in \cite{Kar72} to be NP-hard. Hence, research such as the work by Lemaic and Speckenmeyer in \cite{preprints596} has focused on finding a good algorithm for approximating $\tau(G)$. Most FVS approximation algorithms have three main steps. The first two steps are vertex reduction steps that are repeatedly called until an acyclic digraph is reached. The first step, which is mainly the same across all such FVS algorithms, is to reduce the size of the digraph without changing $\tau(G)$. We will discuss these reductions in section~\ref{sec:digraph-reductions}. The second main step is to choose a vertex in the reduced graph and add it to a candidate FVS. The second step is where most FVS algorithms differ. We discuss various selection schemes in section~\ref{sec:appr-algor-1}. The third main step is post-processing, which assures that the resulting FVS is converted into a minimal one. We will briefly discuss this step in section~\ref{sec:remov-redund-vert}.

\begin{figure}[h!]
  \centering
  \begin{tikzpicture}[main edge/.style = {->,thick},main node/.style={circle,draw,fill=blue!20,thin,inner sep=0pt,minimum size=6mm,font=\sffamily\Large\bfseries}] 
    \node[main node] (z) at (1,-3) {$z$};
    \node[main node] (a) at (-3,-.5) {$a_{1}$};
    \node[main node] (b) at ( -1,-.5) {$b_{1}$};
    \node[main node] (c) at ( -3,1.5) {$c_{1}$};
    \node[main node] (d) at ( -1,1.5) {$d_{1}$};

    \node[main node] (at) at (3,-.5) {$a_{t}$};
    \node[main node] (bt) at ( 5,-.5) {$b_{t}$};
    \node[main node] (ct) at ( 3,1.5) {$c_{t}$};
    \node[main node] (dt) at ( 5,1.5) {$d_{t}$};

    \draw[main edge] (z) to (a);
    \draw[main edge] (z) to [bend left](c);
    \draw[main edge] (z) to (at);
    \draw[main edge] (z) to (ct);

    \draw[main edge] (a) to (b);
    \draw[main edge] (a) to [bend right] (d);

    \draw[main edge] (b) to [bend right] (c);
    \draw[main edge] (b) to (z);

    \draw[main edge] (c) to [bend right] (b);
    \draw[main edge] (c) to (d);

    \draw[main edge] (d) to [bend right] (a);
    \draw[main edge] (d) to  (z);
    
    \draw[main edge] (at) to (bt);
    \draw[main edge] (at) to [bend right] (dt);              

    \draw[main edge] (bt) to [bend right] (ct);
    \draw[main edge] (bt) to (z);

    \draw[main edge] (ct) to [bend right] (bt);
    \draw[main edge] (ct) to (dt);
    
    \draw[main edge] (dt) to [bend right] (at);
    \draw[main edge] (dt) to [bend left] (z);

    \draw [loosely dotted, thick] (-.2,.5) -- (2.2,.5);
  \end{tikzpicture}
  
  \caption{For $t\geq2$ the vertex $z$ is not in a minimum FVS, but is in nearly every DCU and most cycles.}
  \label{fig:counter-Example}
\end{figure}
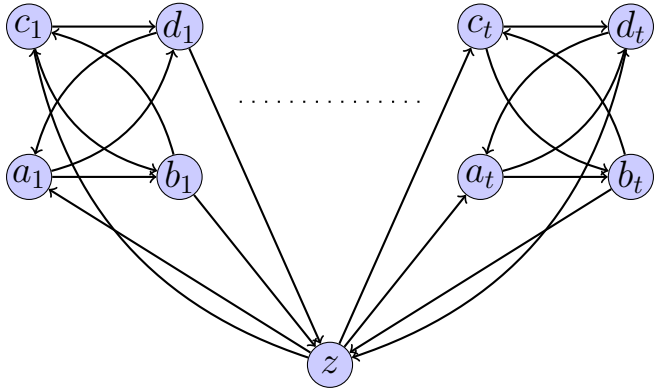

As in \cite{preprints596} we avoid using only local information to choose our FVS candidate. For instance, the authors of \cite{FesParRes01a} and \cite{129914} reasoned that vertices with large out-degrees are more likely to be in many cycles and, hence, are good candidates to be in a minimum FVS. The problem with the maximum degree approach is that it cannot take into account digraphs with many long induced cycles. A digraph composed of vertex disjoint cycles is said to be a \textbf{disjoint cycle union} or DCU. We let $DCU(G)$ be the set of sub-graphs of $G$ that are a DCU. Our method is based on the idea that a vertex that is in many DCUs might also be in a minimum FVS. By the definition of an FVS, every disjoint cycle union of a digraph $G$ has a non-empty vertex intersection with every FVS of $G$. Therefore, for every FVS $S$, there exists an $x\in S$ such that it is in at least $\frac{|DCU(G)|}{|S|}$ DCUs. Thus, if $S$ is a minimum FVS, then some vertex in the set would be in many DCUs. As we see in Figure~\ref{fig:counter-Example} the vertex in the most DCUs or the most simple cycles is not always in a minimum FVS. It does seem reasonable that always choosing a vertex in the most DCUs will yield a good FVS. The problem is that we cannot easily find every DCU of a digraph $G$ as this would be equivalent to calculating the permanent of all the $(n-1)^{2}$ minors of a rank $n$ matrix $\textbf{A}$. Fortunately, Beichl and Sullivan in \cite{Beichl1999128} argued that the Sinkhorn-Knopp algorithm can be used to find a good approximation for the number of spanning DCUs in a digraph. Although Sinkhorn in \cite{Sinkhorn1964} showed that the algorithm would converge under a certain condition and the authors of \cite{Soules1991} and \cite{Knight06thesinkhorn-knopp} gave a rate of convergence, we observed that we did not have to run the algorithm to full convergence. Thus, we present evidence that we can select a good FVS $S$ in about $O(|S|\log(n)n^{2})$ time. If $S$ is an FVS of digraph $G$, then we say $S$ is a $\frac{|S|}{\tau(G)}$-approximation. In Section~\ref{sec:Approximation}, we describe a method that yields a lower bound for $\tau(G)$. We then use this to find a worst-case approximation for our FVS.

Here is an intuitive description of our method. After performing the usual reductions, we add an artificial loop to every vertex, i.e., we put ones along the diagonal of the adjacency matrix. We then Sinkhorn balance the resulting matrix and select the vertex corresponding to the smallest diagonal entry. 

The $i,j^{th}$ entry of the Sinkhorn balanced adjacency matrix approximates the percentage of DCUs that contain the arc from vertex $i$ to vertex $j$. We remove the selected vertex from the digraph and add it to our FVS. We repeat the steps until no cycles remain. This generates an FVS. We finish by removing redundant vertices. These ideas will be made more precise in the rest of the paper.

\begin{center}
\begin{algorithm}[h!]
  \DontPrintSemicolon
  \KwData{A Digraph $G=(V,U)$}
  \KwResult{An FVS $S$}
  \Begin{
    $H \longleftarrow G$\;
    $S \longleftarrow \emptyset$\;
    $LL\_graph\_reductions(H,S)$\;
    $L \longleftarrow get\_SCC(H)$\;
    \While{ $|L| \neq 0$}{
      remove $g$ from $L$\;
      $v \longleftarrow Sinkhorn\_selection(g)$ \;
      remove $v$ from $g$ \;
      $S \longleftarrow S+\{v\}$ \;
      $LL\_reductions(g,S)$ \;	
      $L \longleftarrow get\_SCC(g) + L$\;
    }
    $S \longleftarrow remove\_redundant\_nodes(G,S)$\;
    return $S$
  }
  \caption{FVS\_SH\_Del\label{fvs_sh_del}}
\end{algorithm}
\end{center}
\section{FVS\_SH\_Del Algorithm}
\label{sec:algorithm}
In this section, we explain the components of Algorithm~\ref{fvs_sh_del}. Even though more technical topological sort algorithms exist (See \cite{Ajwani:2012:TSA:2133803.2330083}), we picked a basic topological sort implementation for deciding if a digraph is acyclic.
\subsection{Digraph Reductions}
\label{sec:digraph-reductions}
This subsection deals with reducing the size of a digraph without changing the size of a minimum FVS. Reductions have been studied in detail in \cite{conf/acsc/Koehler05,preprints596, Levy1988470}. We will also use these operations and will briefly list them below.

Clearly, if an arc is not in a cycle we can safely remove it from the digraph. A strongly connected component, denoted by SCC, is a maximal subgraph $H$ such that for any two vertices $u$ and $v$ in $H$ there are directed paths from $u$ to $v$ and from $v$ to $u$. Therefore, we use Tarjan's algorithm, as discussed by Tarjan in \cite{DBLP:journals/siamcomp/Tarjan72}, to find the strongly connected components of the digraph. Any edge not in a strongly connected component is not in a directed cycle of the digraph. Tarjan's algorithm has a worst-case running time of $O(|V|+|E|)$.

We will need the following definition.

\begin{definition}
  \label{def:2}We call the operation of removing a vertex $v$ from a graph $G$ and adding the edges $N^{-}(v)\times N^{+}(v)$ that are not already in $G$ an \textbf{exclusion} of $v$ from $G$. We will follow the notation used in \cite{preprints596} by denoting an exclusion by $G\circ v$.
\end{definition}

Levy and Low in \cite{Levy1988470} listed a few basic operations that can be used to reduce the digraph without changing the FVS problem. We will refer to this operation as LL\_Reductions.

\begin{itemize}
\item $loop(v)$: if there exists a loop, then it is in every FVS and we can safely remove it and add it to our FVS.
\item $in0\_out0(v)$: If $v$ has no successors or predecessors, then $v$ is not in a minimum FVS and we can safely remove it.
\item $in1\_out1(v)$: If $v$ has exactly one successor or one predecessor $u$, then whenever $v$ is in an FVS so is $u$. Thus, we can safely exclude $v$ from $G$. 
\end{itemize}
 
Levy and Low in \cite{Levy1988470} showed that the above reductions can be done in any order. We run an algorithm called LL\_Reductions that recursively checks for Levy-Low reductions until no more reductions can be done. Note that if the Levy-Low reductions return an empty digraph, then a minimum FVS has been found, and we are done. Such digraphs were studied by Koehler in \cite{conf/acsc/Koehler05}.

\subsection{Sinkhorn-Knopp Algorithm}
\label{sec:sinkhorn-balancing}

Matrices will be denoted by bold print. Every digraph $G$ is associated with a $|V| \times |V|$ adjacency matrix $\textbf{A(G)}$. The entry $a_{i,j}$ of $\textbf{A(G)}$ is the number of arcs between $i$ and $j$ in $G$. For a matrix $\textbf{A}$ we will denote $\textbf{A}_{i,j}$ as the minor of $\textbf{A}$ that excludes the row $i$ and column $j$ of $\textbf{A}$. The matrix $\textbf{A}_{i, i}$ is the adjacency matrix of a digraph $G$ with the vertex $i$ removed. A permutation $\sigma$ of a $n\times n$ non-negative matrix $\textbf{A}$ has the property that $a_{i,\sigma(i)}>0$ for all $i$. A matrix is said to be \textbf{fully supported} if every positive entry is in some permutation.
The permanent of a matrix is defined as
\begin{equation}
  \label{eq:1}
  perm(\textbf{A})= \sum_{\sigma}\prod_{i=1}^{n}a_{i,\sigma(i)}.
\end{equation}
A matrix is said to be doubly stochastic if its row and column sums equal one.

Donald, Elwin, Hager, and Salamon in \cite{Donald1984187} showed that the number of spanning disjoint cycle unions, SDCU for short, of a digraph $G$ is equal to the permanent of the adjacency matrix of $G$. For our problem, we need to consider all cycles and, thus, all disjoint cycle unions of a digraph. Thus, if we add loops to every vertex of $G$ to create an auxiliary digraph $H$, then we can see that $perm(\textbf{A}(H))-1=|DCU(G)|$. For $\textbf{A}(H)$, we can create a matrix $m\_bal(\textbf{A}(H))$, called the matrix balance of $\textbf{A}$, if every entry has the form 
\begin{equation}\frac{a_{i,j}\times perm(\textbf{A}(H)_{i,j})}{perm(\textbf{A}(H))}.\end{equation}

We can interpret the entries of $m\_bal(\textbf{A}(H))$ as the fraction of times an arc of $H$ is in a spanning disjoint cycle union. If we were able to efficiently calculate the $m\_bal(\textbf{A}(H))$, then the loop with the smallest value would correspond to the vertex that is in most DCUs of $G$. Unfortunately, we are unable to efficiently calculate the matrix balance, so we must look for ways to estimate the $m\_bal(\textbf{A}(H))$. In \cite{Beichl1999128} the authors argued that the solution of the Sinkhorn-Knopp algorithm is a good estimate for the matrix balance. This is where we focus our attention in the search for an FVS.
\begin{center}
\begin{algorithm}[h!]
\DontPrintSemicolon
\KwData{A Digraph $G=(X,U)$}
\KwResult{A vertex $v$}
\Begin{
$\textbf{A} \longleftarrow$ adjacency matrix of $G$\;
$\textbf{A} \longleftarrow$ add ones to the diagonal of $\textbf{A}$\;

\For{$i\in \{1,...,\lceil \log(n) \rceil\}$}{
        $\textbf{A} \longleftarrow$ normalize the rows of $\textbf{A}$\;
        $\textbf{A} \longleftarrow$ normalize the columns of $\textbf{A}$\;
      }
$v$ is the vertex corresponding to the lowest value on the diagonal of $\textbf{A}$\;
return $v$
	}

\caption{Sinkhorn\_Selection\label{Sinkhornselection}}
\end{algorithm}
\end{center}
As seen in Algorithm~\ref{Sinkhornselection}, the Sinkhorn-Knopp algorithm essentially converts a matrix to a doubly stochastic matrix by repeatedly alternating between normalizing the rows and then normalizing the columns. Each iteration of the Sinkhorn-Knopp algorithm applied to a matrix first divides every row by that row's sum and divides every entry of the row normalized matrix by its column sums. Soules in \cite{Soules1991} showed that the Sinkhorn-Knopp algorithm converges linearly if the matrix has total support. Since we first run Tarjan's algorithm on $G$, we know that every edge of $H$ is in a cycle, and thus, \textbf{A(G)} is totally supported. Since the limiting matrix of Algorithm~\ref{Sinkhornselection} may have irrational values, we may have to stop it after a finite number of steps. Fortunately, we only need to find the vertex with the lowest value along the diagonal in the limiting matrix of the Sinkhorn-Knopp algorithm. We have observed that after $\log(|V(G)|)$ iterations, the ordering of the vertices has settled down and we can select a vertex that is close to the lowest valued vertex in the limit. We did not attempt to optimize the number of iterations needed as we only need to show that this is enough to improve on other algorithms. It would be interesting to see if a constant number of iterations would suffice.
 
If $S$ is the FVS produced by Algorithm~\ref{fvs_sh_del}, then the main body of the algorithm has iterated $|S|$ times. For each iteration, the LL\_Reductions was run once followed by Tarjan's algorithm. These two steps take $O(2|V|+|E|)$ operations. Each main iteration was then finished by performing $\log(|V|)$ Sinkhorn-Knopp iterations. In the worst case, every cell of the adjacency matrix was visited during the Sinkhorn-Knopp iterations. Let $n=|V|$. At the worst Algorithm~\ref{fvs_sh_del} takes \begin{equation}O(n+n^{2}+|S|\log(n)n^{2})=O(|S|\log(n)n^{2}) \end{equation} operations. In practice, we have noted that the algorithm finishes rather quickly. With this observation, we modified Algorithm~\ref{fvs_sh_del} to Algorithm~\ref{fvs_sh_del_remove_SCC} by not reducing the digraph into strongly connected components. We found that Algorithm~\ref{fvs_sh_del_remove_SCC} is considerably quicker than Algorithm~\ref{fvs_sh_del} and only gives a slightly worse FVS. In our analysis, we compare these methods to other existing methods.
\begin{center}
\begin{algorithm}[t]
\DontPrintSemicolon
\KwData{A Digraph $G=(V,U)$}
\KwResult{An FVS $S$}
\Begin{
$H \longleftarrow G$\;
$S \longleftarrow \emptyset$\;
$LL\_graph\_reductions(H,S)$\;
\While{ $|V(H)| \neq 0$}{
	$v \longleftarrow Sinkhorn\_selection(H)$ \;
        remove $v$ from $H$ \;
	$S \longleftarrow S+\{v\}$ \;
	$LL\_reductions(H,S)$ \;	
	}
$S \longleftarrow remove\_redundant\_nodes(G,S)$\;
return $S$
	}

\caption{FVS\_SH\_DEL\_MOD\label{fvs_sh_del_remove_SCC}}
\end{algorithm}
\end{center}
\subsection{Removing Redundant Vertices}
\label{sec:remov-redund-vert}
Once our algorithm selects an FVS it is possible that it is not a minimal one. Thus, the third part of Algorithm~\ref{fvs_sh_del} is to reduce $S$ to a minimal one. As discussed by Hasselman in \cite{hasselman2004efficient}, there are several ways of doing this, but for our case, we picked a simple method. Let $S=S_{0}$ and assume $S_{0}$ is in the reverse order that the vertices of $S$ were selected by Algorithm~\ref{fvs_sh_del}. We then recursively select vertex $v_{i}$ from $S_{i-1}$ and check to see if $G-(S_{i-1}-\{v\})$ is a DAG. If it is not a DAG, then we let $S_{i} = S_{i-1}$. If it is a DAG, then $v$ is redundant and we let $S_{i}=S_{i-1}-\{v\}$.

\section{Analysis}
\label{sec:analysis}

Testing the performance of an FVS algorithm on Erd\H{o}s-Renyi random digraphs and $k$-regular digraphs has been established as a standard for comparing results. Thus, we focus our analysis on these two classes. 

An Erd\H{o}s-Renyi random digraph can be created by visiting every ordered pair of vertices and adding an arc with probability $p$. An $n$ vertex Er\H{o}s-Renyi digraph has expected degree $np$. 

A digraph $G$ is said to be $k$-regular if for every $v\in V(G)$, $d^{+}(v)=d^{-}(v)=k$. We created $k$-regular digraphs by first using a method used by Kleitman and Wang in \cite{Kleitman197379} to generate a $k$-regular digraph. We then repeatedly performed edge switches on that graph to simulate a uniformly selected random $k$-regular digraph. If $(u,v)$ and $(x,y)$ are arcs, $v\not\in N^{+}(x)$, $y \not\in N^{+}(u)$, and $v\neq y$, then deleting $(u,v)$ and $(x,y)$ from $G$ and adding the edges $(u,y)$ and $(x,v)$ is said to be a $2$-switch. Greenhill in \cite{journals/combinatorics/Greenhill11} showed that performing random switching has a polynomial mixing time. The mixing time is too high for practical purposes, and so we performed $k^{2}n$ switches. 
\subsection{Lower Bound}
\label{sec:Approximation}

This section describes how we find a lower bound for $\tau(G)$. A lower bound can be used to show that the Feedback vertex sets we generate improves on the approximation $O(\log(n)\log(\log(n)))$ given by Seymour in \cite{seymourPackingDiCircuitFrac} and improves the bound given in \cite{EvenNSS95}. Recall that an FVS $S$ of a digraph $G$ is a $\frac{|S|}{\tau(G)}$-approximation. We can find an upper bound for this approximation if we find a lower bound for $\tau(G)$. 

We define $\mathcal{C}(G)$ to be the set of cycles of a digraph $G$, and we will let $\mathcal{C}(G,k)$ denote a set of cycles in $\mathcal{C}(G)$ such that every $v\in V(G)$ intersects no more than $k$ cycles in $\mathcal{C}(G,k)$. We say $\mathcal{C}(G,k)$ is \textbf{maximal} if it is not a subset of another $\mathcal{C}^{'}(G,k)$.  For $X\subseteq \mathcal{C}(G)$ we will let $c_{X}(x)$ be the number of cycles in $X$ that hit the vertex $x$ of $G$. A key observation is that if $T$ is an FVS of $G$, then
\begin{equation}\label{E: 1}\sum_{v \in T}{c_{X}(v)}\geq |X|.\end{equation} Let $\alpha$ be an ordering of the vertices of $G$ such that $c_{X}(\alpha_{i})\geq c_{X}(\alpha_{i+1})$ and let $\beta$ be an ordering of the vertices of $T$ such that $c_{X}(\beta_{i})\geq c_{X}(\beta_{i+1})$.  By equation~\ref{E: 1} there exists smallest integers $t$ and $t'$ such that\begin{equation}\sum_{i=0}^{t}{c_{X}(\alpha_{i})}\geq |X| \end{equation} and \begin{equation} \sum_{i=0}^{t'}{c_{X}(\beta_{i})}\geq |X|.\end{equation} By the definition of $\alpha$ and $t$ we have that $t\leq t'\leq \tau(G)$. Thus, we can approximate $\tau(G)$ by $\frac{|S|}{t}$. If $|X|=\epsilon |S|$ and no vertex of $G$ is in more than $k$ cycles of $X$, then we have the bound \begin{equation}k\tau(G)\geq \sum_{i=0}^{t}{c_{X}(\alpha_{i})}\geq |X|=\epsilon |S|.\end{equation} This implies that $|S|$ is a $\frac{k}{\epsilon}$-approximation of $\tau(G)$. Since it is no easy task to find every cycle of a digraph we try to find a subset of cycles $X$ such that no vertex of $G$ is in more than $k$ of them. 

Let $S$ be an FVS of a digraph $G$, and $k=2$. We do the following procedure to get a lower bound on $\tau(G)$ of a digraph $G$. To start we give every vertex of $G$ weight zero. As we find a cycle we add one to the weight of the vertices in the cycle. Let $S=\{s_{0},....,s_{|S|}\}$ be some ordering of $S$, and $G_{0} = G$. For $i$, letting $j = i~mod(|S|)$, we look for a shortest cycle that uses $s_{j}$, but does not intersect $S-\{s_{j}\}$. If no such cycle is found, then let $G_{i+1}=G_{i}$ and proceed with $i+1$. Let $W \subseteq V(C)$ be the set of vertices with weight $k-1$ in $G_{i}$. We add one to the weights of $V(C)$ in $G$ and create a new graph $G_{i+1}$ by deleting the set $W$ and the edges incident with $s_{j}$ along $C$ from $G_{i}$. We stop if $G_{i+1}$ is acyclic or if every cycle of $G_{i+1}$ uses at least two vertices of $S$.

We actually run the above procedure twice. We first run the algorithm once on an FVS $S$ and then let $S'$ be the set of vertices in $S$ that have non-zero weight. If $S=S'$, then we are done. Otherwise, we run the algorithm on $S'$. We have observed that doing this results in a substantial improvement in the number of cycles.

\subsection{Approximation Algorithms}
\label{sec:appr-algor-1}

During the vertex selection phase, many FVS algorithms make selections based on degrees. For instance, some methods always remove the vertex with maximum out-degree. These methods, although quick, have been shown to yield a larger FVS when compared to more sophisticated methods given in \cite{preprints596}. Algorithm~\ref{MaxDeg} simply removes the vertex $v$ that satisfies $min(d^{+}(v),d^{-}(v))\geq min(d^{+}(u),d^{-}(u))$ for all $u\in V(G)$. 

\begin{algorithm}[t]
\DontPrintSemicolon
\KwData{A Digraph $G=(V,U)$}
\KwResult{An FVS $S$}
\Begin{
$S \longleftarrow \emptyset$\;
$LL\_graph\_reductions(G,S)$\;
$L \longleftarrow get\_SCC(G)$\;
\While{ $|L| \neq 0$}{
	remove $g$ from $L$\;
	$v \longleftarrow max(min(d^{+}(v),d^{-}(v))|v\in V(G))$ \;
	remove $v$ from $g$ \;
	$S \longleftarrow S+\{v\}$ \;
	$LL\_reductions(g,S)$ \;	
	$L \longleftarrow get\_SCC(g) + L$\;
	}
$S \longleftarrow remove\_redundant\_nodes(G,S)$\;
return $S$
	}

\caption{MaxDeg \label{MaxDeg}}
\end{algorithm}

The authors of \cite{speck-Feedbackproblems} and of \cite{preprints596} studied a random walk on a digraph $G$. They argued that the vertex with the smallest mean return time is probably in many small cycles. Small cycles are interesting because the probability that a randomly selected vertex on a cycle $C$ is in a minimum FVS of $G$ is $\frac{1}{|C|}$. Thus, it is reasonable to suspect that a vertex $v$ that is in many small cycles is in a minimum FVS of $G$. A transition matrix $\textbf{P}$ can be created from the adjacency matrix $\textbf{A(G)}$ by normalizing the rows. If $\pi$ is the stationary distribution of $\textbf{P}$, then $\frac{1}{\pi_{v}}$ is equal to the mean return time for $v$. We let $G^{-1}$ be the digraph in which the arcs of $G$ are reversed. The authors made another observation that if they took the sum of the stationary distributions of $G$ and $G^{-1}$, then a better FVS may be obtained. Thus, we have Algorithm~\ref{MFVSmean}. Their algorithm can be done theoretically in $O(|F|n^{2.376})$ operations. They gave an in-depth discussion of how their algorithm improves on other known methods. We will present evidence that suggests Algorithm~\ref{fvs_sh_del} produces a smaller solution than Algorithm~\ref{MFVSmean} in most cases.

\begin{algorithm}[t]
\DontPrintSemicolon
\KwData{A Digraph $G=(V,U)$}
\KwResult{A vertex $v$}
\Begin{
$\textbf{P} \longleftarrow CreateTransitionMatrix(G)$\;
$\pi' \longleftarrow ComputeStationaryDistributionVector(\textbf{P})$\;
$\textbf{P} \longleftarrow CreateTransitionMatrix(G^{-1})$\;
$\pi'' \longleftarrow ComputeStationaryDistributionVector(\textbf{P})$\;

$\pi \longleftarrow \pi' + \pi''$\;

determine $v\in V$ with $\pi_{v} = \|\pi\|_{\infty}$\;
return $v$
	}

\caption{MFVSmean\_selection\label{MFVSmeanselection}}
\end{algorithm}

\begin{algorithm}[h!]
\DontPrintSemicolon
\KwData{A Digraph $G=(V,U)$}
\KwResult{An FVS $S$}
\Begin{
$S \longleftarrow \emptyset$\;
$LL\_graph\_reductions(G,S)$\;
$L \longleftarrow get\_SCC(G)$\;
\While{ $|L| \neq 0$}{
	remove $g$ from $L$\;
	$v \longleftarrow MFVSmean\_selection(g)$ \;
	remove $v$ from $g$ \;
	$S \longleftarrow S+\{v\}$ \;
	$LL\_reductions(g,S)$ \;	
	$L \longleftarrow get\_SCC(g) + L$\;
	}
$S \longleftarrow remove\_redundant\_nodes(G,S)$\;
return $S$
	}

\caption{MFVSmean\label{MFVSmean}}
\end{algorithm}

\subsection{Erd\H{o}s-Renyi Random Digraphs}
\label{sec:random-digraphs}

We created Erd\H{o}s-Renyi random digraphs by visiting every ordered pair of vertices and placing an arc with probability $p$ between them. At least one hundred digraphs were created for every $n$ and $p$ tested. We tested the various FVS algorithms on each digraph we created with a given $n$ and $p$. For each digraph $G$, we used the approximation method to find a lower bound for $\tau(G)$ and calculated the approximation each method is guaranteed to achieve. In the pictures below, we plotted the mean approximation for all digraphs with a fixed $p$, $n$, and method.

\begin{figure}[h!]
  \centering
  \includegraphics[scale = 0.5]{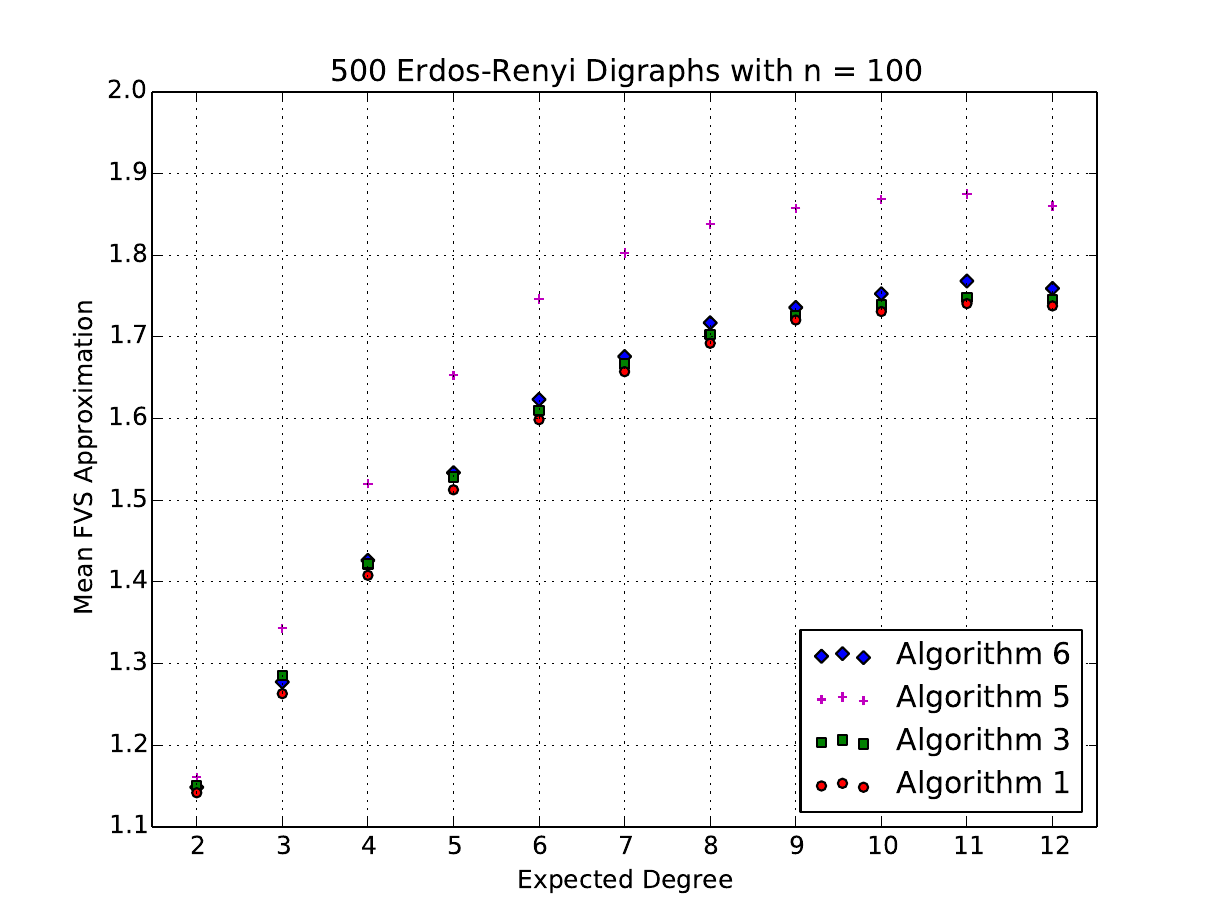}
  \caption{The $x$-axis is the expected degree and the $y$-axis is the worst case mean approximation for each algorithm on $500$ Erd\H{o}s-Renyi digraphs with $100$ vertices.}
  \label{fig:ER_100}
\end{figure}

\begin{figure}[h!]
  \centering
  \includegraphics[scale = 0.5]{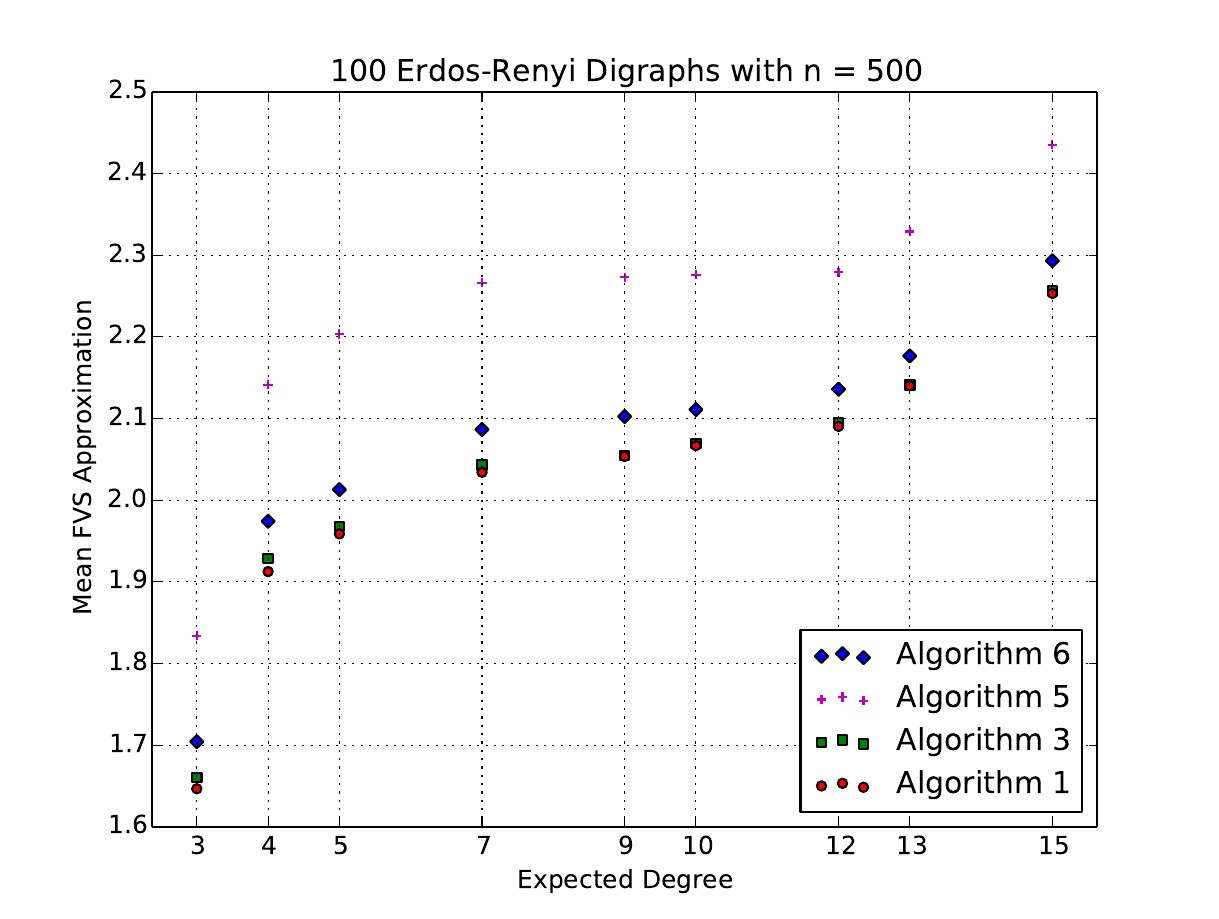}
  \caption{The $x$-axis is the expected degree and the $y$-axis is the worst case mean approximation for each algorithm on $100$ Erd\H{o}s-Renyi digraphs with $500$ vertices.}
  \label{fig:ER_500}
\end{figure}

\begin{figure}[h!]
  \centering
  \includegraphics[scale = 0.5]{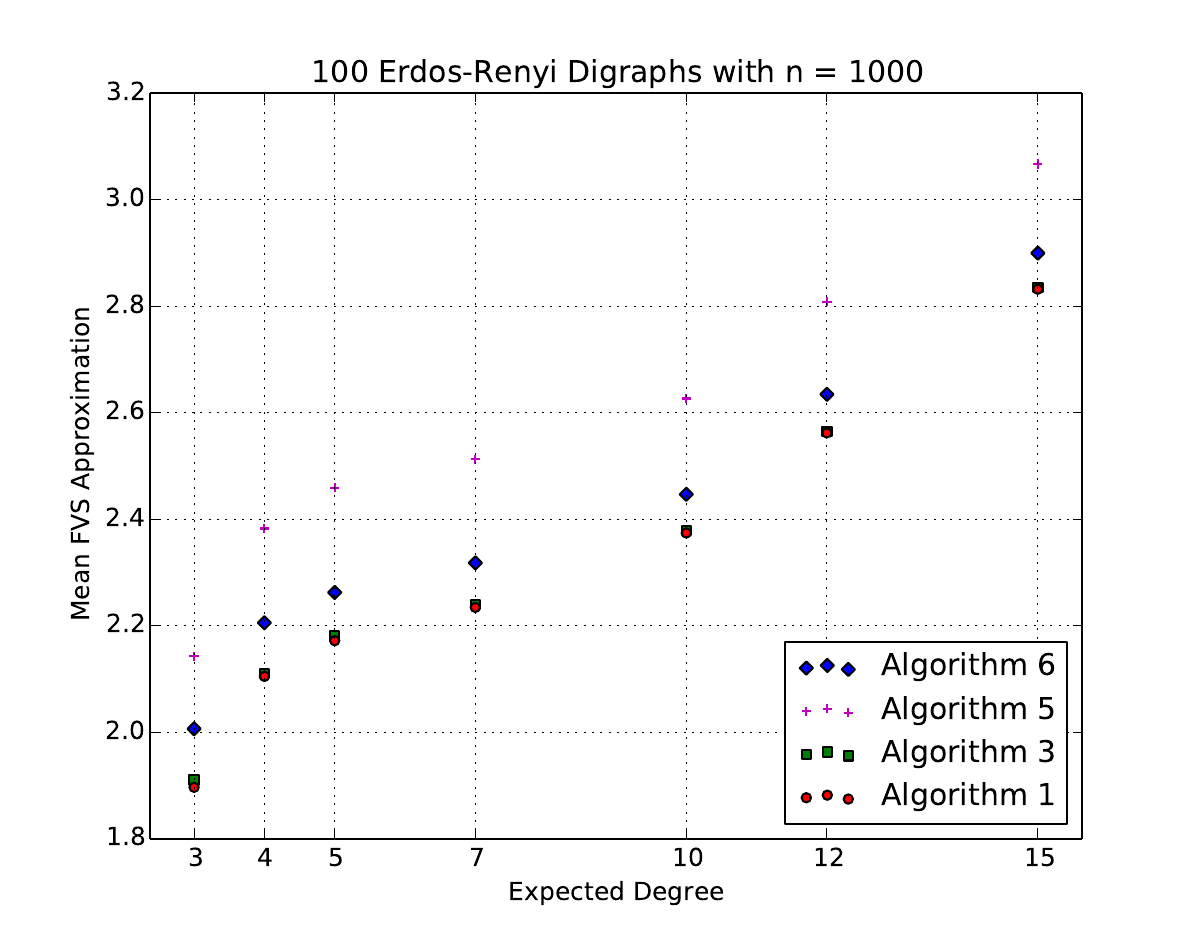}
  \caption{The $x$-axis is the expected degree and the $y$-axis is the worst case mean approximation for each algorithm on $100$ Erd\H{o}s-Renyi digraphs with $1000$ vertices.}
  \label{fig:ER_1000}
\end{figure}

In Figure~\ref{fig:ER_100}, Algorithm~\ref{fvs_sh_del} clearly performs on average better than the others. What is interesting is that Algorithm~\ref{fvs_sh_del_remove_SCC} performs better than Algorithm~\ref{MaxDeg} and Algorithm~\ref{MFVSmean}. In Figure~\ref{fig:ER_500} and Figure~\ref{fig:ER_1000} we can see that the worst-case mean approximation of Algorithm~\ref{fvs_sh_del_remove_SCC} is very close to Algorithm~\ref{fvs_sh_del} and performs better than Algorithm~\ref{MaxDeg} and Algorithm~\ref{MFVSmean} for all $p$. In practice, Algorithm~\ref{fvs_sh_del_remove_SCC} is considerably faster than all the others. This suggests that more time is spent on Tarjan's algorithm in Algorithm~\ref{fvs_sh_del} than the scaling steps.

\subsection{Regular digraphs}
\label{sec:regular-digraphs}

One hundred $k$-regular digraphs were created for every $n$ and $k$ tested. For each digraph $G$, we used the approximation method to find a lower bound for $\tau(G)$ and calculated the approximation each method is guaranteed to achieve. In the pictures below, we plotted the mean approximation for all digraphs with a fixed $n$, $k$, and for each method.

\begin{figure}[h!]
  \centering
  \includegraphics[scale = 0.5]{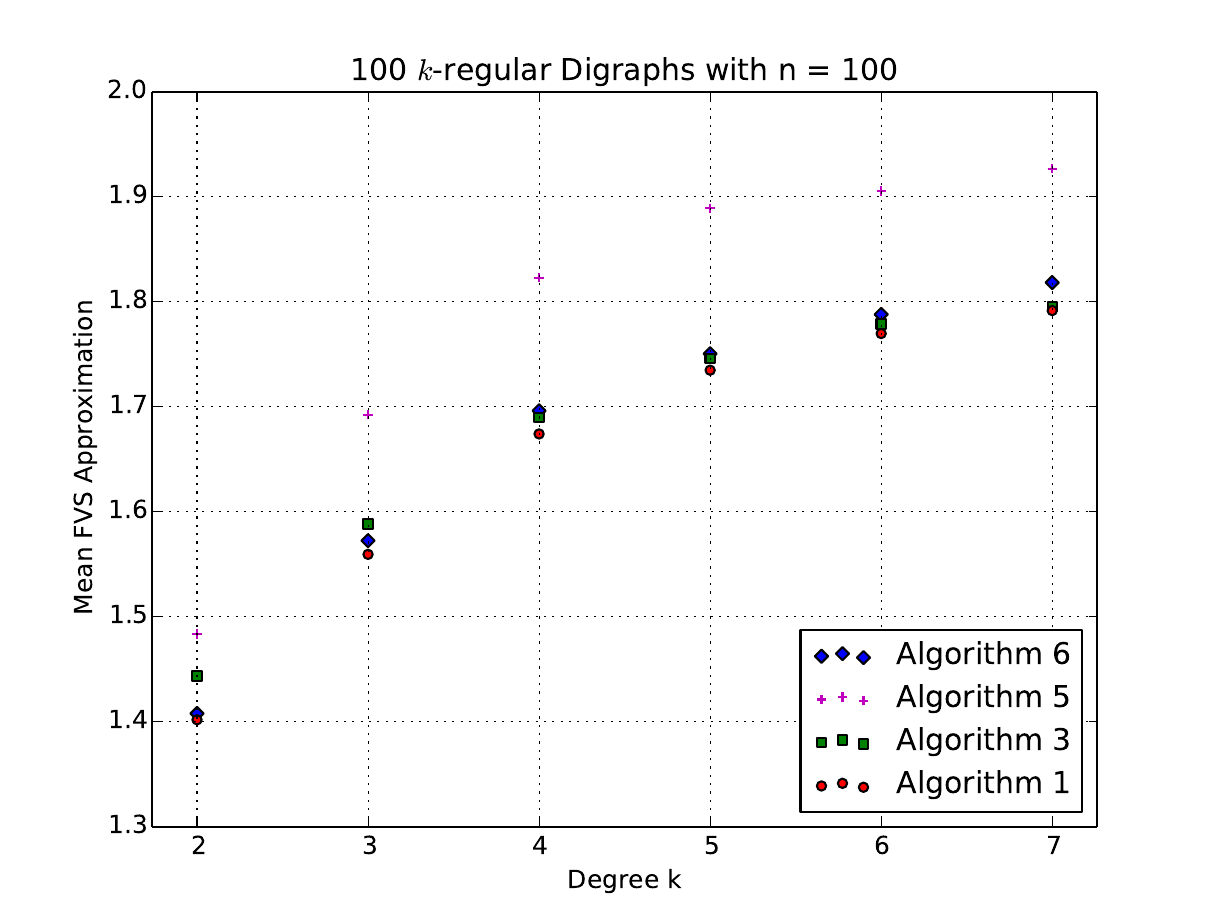}
   \caption{The $x$-axis is the degree and the $y$-axis is the worst case mean approximation for each algorithm on $100$ $k$-regular digraphs with $100$ vertices.}
  \label{fig:RK_100}
\end{figure}

Figure~\ref{fig:RK_100} shows that for sparse digraphs Algorithm~\ref{fvs_sh_del} still performs best, but Algorithm~\ref{fvs_sh_del_remove_SCC} does not perform better than Algorithm~\ref{MFVSmean} until the $4$-regular digraphs. Algorithm~\ref{fvs_sh_del_remove_SCC} did not produce smaller sets than Algorithm~\ref{fvs_sh_del}, but in practice Algorithm~\ref{fvs_sh_del_remove_SCC} finished significantly faster than other methods.
\begin{figure}[h!]
  \centering
  \includegraphics[scale = 0.5]{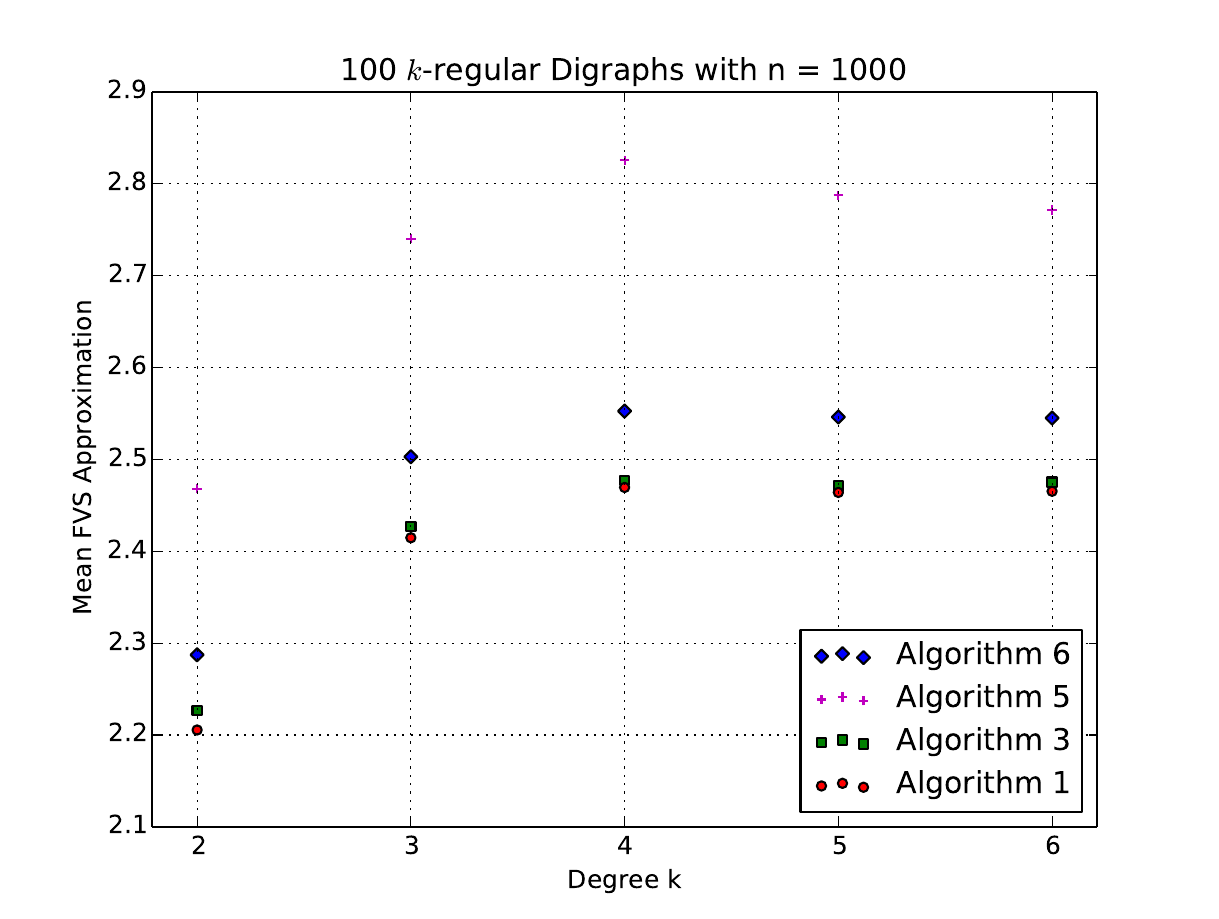}
  \caption{The $x$-axis is the degree and the $y$-axis is the worst case mean approximation for each algorithm on $100$ $k$-regular digraphs with $1000$ vertices.}

 \label{fig:RK_1000}
\end{figure}

Figure~\ref{fig:RK_1000} shows that both Algorithm~\ref{fvs_sh_del} and Algorithm~\ref{fvs_sh_del_remove_SCC} perform better than the other methods on $k$-regular digraphs with $1000$ nodes.

\section{Observation}
\label{sec: observation}

For completion, we compared Algorithm~\ref{fvs_sh_del} to one where we used the $m\_$balance for the vertex selections. Due to the complexity of this we were only able to compare digraphs with twenty-five nodes or fewer. Surprisingly, most of the time we found that the Sinkhorn-Knopp balance selection performed equal to or better than the modified algorithm. This indicates that the DCUs play a part, but there is something more happening with the Sinkhorn-Knopp algorithm. Perhaps it has something to do with entropy. Recall that the entropy of a doubly stochastic matrix \textbf{A} is defined as  \begin{equation}
  \label{eq:2}
 entropy(\textbf{A})= -\sum_{i,j}a_{i,j}\log(a_{i,j}).
\end{equation}
Beichl and Sullivan in \cite{Beichl1999128} showed the limiting matrix of the  Sinkhorn-Knopp algorithm maximizes the entropy for all doubly-stochastic matrices with a given zero-one pattern. Perhaps one could fix a digraph and generate different doubly stochastic matrices to see if there is any correlation.

\section{Conclusion}
\label{sec:conclusions}

We have presented empirical evidence that Algorithm~\ref{fvs_sh_del} produces smaller feedback vertex sets than other algorithms found in the literature. We noted that our algorithm is faster than the worst-case complexity for Algorithm~\ref{MFVSmean}. We found with Algorithm~\ref{fvs_sh_del_remove_SCC} that not reducing the digraph into strongly connected components improves the complexity of Algorithm~\ref{fvs_sh_del} without affecting the quality of the answer significantly. Finally, we give empirical evidence that our algorithm produces an approximation of an optimal FVS that is well below the average degree of the digraph. 

\section*{Acknowledgment}
We want to thank Francis Sullivan for his helpful discussions.


\begin{thebibliography}{10}

\bibitem{Ajwani:2012:TSA:2133803.2330083}
Deepak Ajwani, Adan Cosgaya-Lozano, and Norbert Zeh, \emph{A topological sorting algorithm for large graphs}, J. Exp. Algorithmics \textbf{17} (2012), 3.2:3.1--3.2:3.21.

\bibitem{Bathie2022}
Gabriel Bathie, Ga{\'e}tan Berthe, Yoann Coudert-Osmont, David Desobry, Amadeus Reinald, and Mathis Rocton, \emph{{PACE solver description: DreyFVS}}, {Leibniz International Proceedings in Informatics } (2022).

\bibitem{Beichl1999128}
Isabel Beichl and Francis Sullivan, \emph{Approximating the permanent via importance sampling with application to the dimer covering problem}, Journal of Computational Physics \textbf{149} (1999), no.~1, 128 -- 147.

\bibitem{DBLP:books/daglib/0030488}
Reinhard Diestel, \emph{Graph theory, 4th edition}, Graduate texts in mathematics, vol. 173, Springer, 2012.

\bibitem{Donald1984187}
John Donald, John Elwin, Richard Hager, and Peter Salamon, \emph{A graph theoretic upper bound on the permanent of a nonnegative integer matrix. i}, Linear Algebra and its Applications \textbf{61} (1984), no.~0, 187 -- 198.

\bibitem{EvenNSS95}
Guy Even, Joseph Naor, Baruch Schieber, and Madhu Sudan, \emph{Approximating minimum feedback sets and multi-cuts in directed graphs.}, IPCO (Egon Balas and Jens Clausen, eds.), Lecture Notes in Computer Science, vol. 920, Springer, 1995, pp.~14--28.

\bibitem{FesParRes01a}
P.~Festa, P.M. Pardalos, and M.G.C. Resende, \emph{Algorithm 815: {FORTRAN} subroutines for computing approximate solution to feedback set problems using {GRASP}}, ACM Transactions on Mathematical Software \textbf{27} (2001), 456--464.

\bibitem{journals/combinatorics/Greenhill11}
Catherine~S. Greenhill, \emph{A polynomial bound on the mixing time of a markov chain for sampling regular directed graphs.}, Electr. J. Comb. \textbf{18} (2011), no.~1.

\bibitem{hasselman2004efficient}
B.H. Hasselman, \emph{An efficient method for detecting redundant feedback vertices}, CPB discussion paper, CPB Netherlands Bureau for Economic Policy Analysis, 2004.

\bibitem{Kar72}
R.~Karp, \emph{Reducibility among combinatorial problems}, Complexity of Computer Computations (R.~Miller and J.~Thatcher, eds.), Plenum Press, 1972, pp.~85--103.

\bibitem{Kleitman197379}
D.J. Kleitman and D.L. Wang, \emph{Algorithms for constructing graphs and digraphs with given valences and factors}, Discrete Mathematics \textbf{6} (1973), no.~1, 79 -- 88.

\bibitem{Knight06thesinkhorn-knopp}
Philip~A. Knight, \emph{The sinkhorn-knopp algorithm: convergence and applications}, Tech. report, 2006.

\bibitem{conf/acsc/Koehler05}
Henning Koehler, \emph{A contraction algorithm for finding minimal feedback sets}, ACSC (Vladimir Estivill-Castro, ed.), CRPIT, vol.~38, Australian Computer Society, 2005, pp.~165--174.

\bibitem{129914}
D.H. Lee and S.M. Reddy, \emph{On determining scan flip-flops in partial-scan designs}, Computer-Aided Design, 1990. ICCAD-90. Digest of Technical Papers., 1990 IEEE International Conference on, 1990, pp.~322--325.

\bibitem{preprints596}
Mile Lemaic and Ewald Speckenmeyer, \emph{Markov-chain-based heuristics for the minimum feedback vertex set problem}, Technical report, 2009.

\bibitem{Levy1988470}
Hanoch Levy and David~W Low, \emph{A contraction algorithm for finding small cycle cutsets}, Journal of Algorithms \textbf{9} (1988), no.~4, 470 -- 493.

\bibitem{seymourPackingDiCircuitFrac}
P.~D. Seymour, \emph{Packing directed circuits fractionally}, Combinatorica \textbf{15} (1995), no.~2, 281--288.

\bibitem{Shook2014}
James Shook and Isabel Beichl, \emph{Matrix scaling: a new heuristic for the feedback vertex set problem}, https://math.nist.gov/mcsd/Seminars/2014/2014-06-10-Shook-presentation.pdf, June 2014.

\bibitem{Sinkhorn1964}
Richard Sinkhorn, \emph{A relationship between arbitrary positive matrices and doubly stochastic matrices}, The Annals of Mathematical Statistics \textbf{35} (1964), no.~2, 876--879.

\bibitem{Soules1991}
George~W. Soules, \emph{{The rate of convergence of Sinkhorn balancing}}, Linear Algebra and its Applications \textbf{150} (1991), 3--40.

\bibitem{speck-Feedbackproblems}
Ewald Speckenmeyer, \emph{On feedback problems in digraphs}, Graph-Theoretic Concepts in Computer Science (Manfred Nagl, ed.), Lecture Notes in Computer Science, vol. 411, Springer Berlin Heidelberg, 1990, pp.~218--231.

\bibitem{DBLP:journals/siamcomp/Tarjan72}
Robert~Endre Tarjan, \emph{Depth-first search and linear graph algorithms}, SIAM J. Comput. \textbf{1} (1972), no.~2, 146--160.

\end{thebibliography}

\providecommand{\bysame}{\leavevmode\hbox to3em{\hrulefill}\thinspace}
\providecommand{\MR}{\relax\ifhmode\unskip\space\fi MR }
\providecommand{\MRhref}[2]{%
  \href{http://www.ams.org/mathscinet-getitem?mr=#1}{#2}
}
\providecommand{\href}[2]{#2}

\end{document}